# Variational Stochastic Games

Zhiyu Zhao
The Institute of Automation of the Chinese Academy of Sciences
Beijing, China
School of Artificial Intelligence
University of the Chinese Academy of Sciences
Beijing, China
zhaozhiyu2022@ia.ac.cn

Haifeng Zhang
The Institute of Automation of the Chinese Academy of Sciences
Beijing, China
School of Artificial Intelligence
University of the Chinese Academy of Sciences
Beijing, China
haifeng.zhang@ia.ac.cn

## Abstract

The Control as Inference (CAI) framework has successfully transformed single-agent reinforcement learning (RL) by reframing control tasks as probabilistic inference problems. However, the extension of CAI to multi-agent, general-sum stochastic games (SGs) remains underexplored, particularly in decentralized settings where agents operate independently without centralized coordination. In this paper, we propose a novel variational inference framework tailored to decentralized multi-agent systems. Our framework addresses the challenges posed by non-stationarity and unaligned agent objectives, proving that the resulting policies form an $\epsilon$-Nash equilibrium. Additionally, we demonstrate theoretical convergence guarantees for the proposed decentralized algorithms. Leveraging this framework, we instantiate multiple algorithms to solve for Nash equilibrium, mean-field Nash equilibrium, and correlated equilibrium, with rigorous theoretical convergence analysis.

## CCS Concepts

• **Computing methodologies** → **Multi-agent planning**.

## Keywords

Variational Inference, Stochastic Games, Nash Equilibrium, Mean Field Game, Correlated Equilibrium

## 1 Introduction

Casting a control problem as an inference problem has a long history dating back to the work solving optimal control in linear systems [10]. There have been works framing the problem of reinforcement learning (RL) in the language of variational inference [1, 11, 12, 25]. Agents infer actions that maximize the likelihood of optimal trajectories instead of maximizing the long-term return.

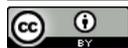

This reformulation enables the use of powerful variational inference tools in RL. The variational inference algorithm can handle the uncertainties regarding the transition probabilities in the environment and provide a natural exploration strategy based on entropy maximization, which improves the stability of the learned policy in complex stochastic systems [6]. Although control as inference (CAI) has led to a series of impressive successes in single-agent tasks, there are limited works that generalize CAI to multi-agent decision-making problems. The multi-agent decision-making problems are often modeled under the framework of general-sum stochastic game (SG). In an SG, each agent interacts with the environment and other agents to maximize its own long-term return. Compared with the single agent tasks, the decision-making problem in the SG has two key challenges [29]. Since all the agent improves their policies to optimize their own objectives simultaneously, the environment is non-stationary in the view of each agent. The non-stationary property increases the uncertainty in the environment. Furthermore, the objectives of agents are multi-dimensional. The objectives of different agents are not necessarily aligned. Hence agents should consider other agents' responses while improving their policies according to their objectives, which requires agents to have the ability to reason about others' behaviors [8, 16, 27].

Although there exists work trying to address the above challenges under the framework of variational inference [21], which limits the scope within the cooperative setting, where agents collaborate to optimize a common long-term return. This assumption is quite restrictive as it is not suitable for the competitive or mixed cooperative-competitive game. In addition, the equivalence between the solution of probabilistic inference and the equilibrium has not been rigorously studied.

In this paper, we aim to propose a unified decentralized variational inference framework to solve general-sum SG. Our main contribution is to reformulate and solve the general-sum stochastic game under the framework the variational inference, which bridges the variational inference and finding the equilibrium in the general-sum stochastic game. This approach allows for the use of more tools to solve the general-sum stochastic game. We redefine the binary random variable optimality to formulate the SG with variational inference. Then we propose a unified variational inference framework for solving general-sum SG. We derive different algorithms under this framework, for solving different equilibrium concepts.

Our contributions are as follows:

- We introduce a unified, decentralized variational inference framework for solving general-sum games. Compared with previous work [21, 28], our framework can handle a wider class of games, whether it is cooperative or not. This framework builds the connection between inference and game theory, allowing recent advances in statistical inference to be applied to diverse sets of game theoretical settings.
- We rigorously prove that the policies derived from our framework form an $\epsilon$-Nash equilibrium and provide formal convergence guarantees for the decentralized algorithms, which is a significant step towards understanding the relationship between variational inference and equilibrium concepts.
- We propose a novel opponent modeling approach based on variational inference, enabling agents to infer the strategies of others even when their objectives are unaligned.
- We instantiate several decentralized algorithms under this framework to solve various equilibrium concepts, including Nash equilibrium, mean-field Nash equilibrium, and correlated equilibrium.

This decentralized approach introduces significant advantages in scalability and flexibility, making it suitable for large-scale multi-agent systems.

## 2 Related works

Applying probabilistic inference to control has a long history [17, 18, 23–25]. Casting a control problem into a probability inference problem enables the application of advanced inference tools to the control, and extends the model of control [11]. However, most of the existing works focus on the single-agent case. There are a few works that try to extend the inference framework to the multi-agent setting. And most of them focus on cooperative games [22, 28], which limits the application of the framework. In our work, we establish a novel variational inference framework for a general-sum stochastic game. We further show that solving mean field games, zero-sum games, and correlated equilibrium in the stochastic game are special cases of the general framework. In addition, previous works introduce a binary random variable named optimality to cast the problem of finding the equilibrium in the multi-agent cooperative game to a probability inference problem. However, the objective of the probability inference problem is different from the objective of finding an equilibrium. The relation between the solution of the probability inference problem and the solution of the original problem is not illustrated. In our work, we find that the solution of probability inference is an $\epsilon$-Nash equilibrium by deriving the performance difference between two solutions.

There are also a few works that apply entropy regularized reinforcement learning method to game theory [7, 26]. However, they treat the entropy or KL divergence as a heuristic modification to enable improved exploration and convergence. In our work, we first formulate the problem under the variational inference framework and derive the entropy or KL divergence naturally in theory.

In this work, we give a new definition of optimality in the general-sum stochastic game and propose a method for solving general-sum stochastic game under the framework of variational inference, which incorporates the opponent modeling. We further apply the framework to various games to show generality.

## 3 Preliminaries
### 3.1 Stochastic game

We consider a SG [19, 20] with $N$ players. The horizon of SG is $\mathcal{T} = \{0, 1, \cdots, T\}$. At each time index $t \in \mathcal{T}$, agent $i \in \mathcal{N}$ ($\mathcal{N} = \{1, 2, \cdots, N\}$) at state $s_t \in \mathcal{S}$ will select an action $a_t^i$ from the action space $\mathcal{A}^i$. All the agents take action simultaneously. Let $\boldsymbol{a}_t = (a_t^1, a_t^2, \cdots, a_t^N) \in \mathcal{A}$ denote the joint action. Each agent will receive a reward $r^i(s_t, \boldsymbol{a}_t)$ and the joint state will change to $s_{t+1}$ according to the transition kernel $P(s_{t+1}|s_t, \boldsymbol{a}_t)$. Agents take actions according to some policy $\pi_i : \mathcal{S} \to \mathcal{P}(\mathcal{A}^i)$. Given the joint policy $\boldsymbol{\pi} = (\pi_1, \pi_2, \cdots, \pi_N)$, the cumulative reward of agent $i$ is

$$V^i(s; \boldsymbol{\pi}) = \sum_{t=0}^{T} \mathbb{E}\left[r_t^i(s_t, \boldsymbol{a}_t)|s_0 = s, \boldsymbol{\pi}\right],$$

where the expectation is taken with respect to $s_{t+1} \sim P(\cdot|s_t, \boldsymbol{a}_t)$, $\boldsymbol{a}_t \sim \boldsymbol{\pi}(\cdot|\boldsymbol{s}_t)$. The Nash equilibrium is a joint policy $\boldsymbol{\pi}^* = (\pi_1^*, \pi_2^*, \cdots, \pi_N^*)$ such that for all agent $i$,

$$V^i(s; \boldsymbol{\pi}^*) \geq V^i(s; \pi_i, \boldsymbol{\pi}_{-i}^*),$$

where $\boldsymbol{\pi}_{-i}^* = (\pi_1^*, \cdots, \pi_{i-1}^*, \pi_{i+1}^*, \cdots, \pi_N^*)$, i.e. $\pi_i^*$ is the best response of $\boldsymbol{\pi}_{-i}^*$. Accordingly, $\pi_i^* \in \Pi_i := \Delta(\mathcal{A}_i)^\mathcal{S}$, $\boldsymbol{\pi}^* \in \Pi := \times_{i \in \mathcal{N}} \mathcal{A}_i$ and $\boldsymbol{\pi}_{-i}^* \in \Pi := \times_{i \neq j \in \mathcal{N}} \mathcal{A}_j$. Similarly, a joint policy $\boldsymbol{\pi}^*$ is the $\epsilon$-Nash equilibrium if there exists an $\epsilon > 0$ so that for all agent $i \in \mathcal{N}$,

$$V^i(s; \boldsymbol{\pi}^*) \geq \max_{\pi_i \in \Pi_i} V^i(s; \pi_i, \boldsymbol{\pi}_{-i}^*) - \epsilon.$$

### 3.2 Control as inference

Control as inference [11] incorporates the reward function by introducing a binary optimality variable $O_t^i$. $O_t^i$ indicates "optimality" for each agent $i$ at each time step $t$. The action of agent $i$ is optimal if $O_t^i = 1$. Maximizing the long-term return in the framework of control is equivalent to maximizing the probability of $O_t^i = 1$. Therefore, a control problem can be cast as an inference problem.

## 4 Variational Stochastic Game: Theory

In this section, we establish a general framework of variational inference for solving stochastic games. We introduce the definition of optimality in the general-sum stochastic game, formalizing the general-sum stochastic game as probabilistic inference. We further establish the relationship between the Nash equilibrium condition and optimality in inference.

To solve the stochastic game using variational inference method, we first build the graphical model of stochastic game as shown in Figure 1. We introduce the variable optimality $O_t^i$ to indicate whether the agent $i$ achieves optimality at time step $t$ in the stochastic game.

**Definition 4.1.** Optimality $O_t^i$ is a binary variable indicating "optimality" for each agent $i$ at each time step $t$. If the optimality for agent $i$ at time step $t$ is reached, the binary variable $O_t^i = 1$. And we have $P(O_t^i = 1|s, \boldsymbol{a}, \boldsymbol{\pi}_{-i}) \propto \exp(r^i(s, \boldsymbol{a}))$.

The objective of the inference problem is to maximize the likelihood $P(O_{0:T}^i = 1|\pi_i, \boldsymbol{\pi}_{-i})$ for all $i \in \mathcal{N}$.

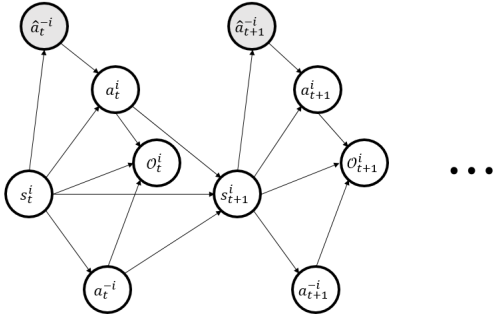

Figure 1: We build the directed graphical model of a stochastic game. The arrows show the conditional dependencies among the variables. The state $s_{t+1}$ and optimality $O_t^i$ for each agent are dependent on $a_t^i$, $a_t^{-i}$, and $s_t$. At each time step, all the agents take action simultaneously. As agent $i$ cannot observe $a_t^{-i}$ in advance, it will use $\hat{a}_t^{-i} \sim \rho(\cdot|s_t)$ (the gray node) to predict actions of other agents. And action $a_t^i$ is dependent on $\hat{a}_t^{-i}$.

Under the framework of variational inference, we transform the objective of agent $i$ to choose a policy to maximize the log-likelihood of optimality.

$$\log P(O_{0:T}^i = 1|\pi_i, \boldsymbol{\pi}_{-i}) = \\ \log \sum_{a_{0:T}^i, a_{0:T}^{-i}, s_{0:T}} P(O_{0:T}^i = 1, a_{0:T}^i, a_{0:T}^{-i}, s_{0:T}|\pi_i, \boldsymbol{\pi}_{-i}) \quad (1)$$

In order to optimize the objective in the (1), we factorize $P(O_{0:T}^i = 1, a_{0:T}^i, a_{0:T}^{-i}, s_{0:T}|\pi_i, \boldsymbol{\pi}_{-i})$ based on the Figure 1. We use an auxiliary distribution over states and actions $q(a_{0:T}^i, a_{0:T}^{-i}, s_{0:T})$ to handle the unknown the transition dynamic and opponent policy.

$$q(a_{0:T}^i, a_{0:T}^{-i}, s_{0:T}) \\ = P(s_0) \prod_{t=0}^{T} q(a_t^i|s_t, a_t^{-i}) q(a_t^{-i}|s_t) P(s_{t+1}|s_t, a_t^i, a_t^{-i}) \\ = P(s_0) \prod_{t=0}^{T} \pi_i(a_t^i|s_t, a_t^{-i}) \rho(a_t^{-i}|s_t) P(s_{t+1}|s_t, a_t^i, a_t^{-i})$$

The opponent policy $\boldsymbol{\pi}_{-i}$ is approximated by agent $i$'s opponent model $\rho(a_t^{-i}|s_t) = \prod_{j \in \mathcal{N}, j \neq i} \rho_j(a_t^j|s_t)$. We can derive the lower bound of (1).

$$\log P(O_{0:T}^i = 1|\pi_i, \boldsymbol{\pi}_{-i}) \\ = \log \sum_{a_{0:T}, s_{0:T}} P(O_{0:T}^i = 1, a_{0:T}, s_{0:T}|\pi_i, \boldsymbol{\pi}_{-i}) \\ = \log \sum_{a_{0:T}, s_{0:T}} q(a_{0:T}, s_{0:T}) \frac{P(O_{0:T}^i = 1, a_{0:T}, s_{0:T}|\pi_i, \boldsymbol{\pi}_{-i})}{q(a_{0:T}, s_{0:T})} \\ \geq \sum_{a_{0:T}, s_{0:T}} q(a_{0:T}, s_{0:T}) \log \frac{P(o_{0:T}^i = 1, a_{0:T}, s_{0:T}|\pi_i, \boldsymbol{\pi}_{-i})}{q(a_{0:T}, s_{0:T})} \quad (2) \\ = \sum_{t=0}^{T} \mathbb{E}[r_t^i(s_t, a_t)] + H(\pi_i(a_t^i|s_t, a_t^{-i})) \\ - \text{KL}(\rho(a_t^{-i}|s_t)||\boldsymbol{\pi}_{-i}(a_t^{-i}|s_t))|\pi_i, \boldsymbol{\pi}_{-i}],$$

where the expectation is taken with respect to $a_t^i \sim \pi_i(\cdot|s_t, a_t^{-i})$, $a_t^{-i} \sim \rho(\cdot|s_t)$, $s_{t+1} \sim P(\cdot|s_t, a_t^i, a_t^{-i})$ and we use the fact that $P(O_t^i = 1|s, \boldsymbol{a}, \boldsymbol{\pi}_{-i}) \propto \exp(r^i(s, \boldsymbol{a}))$.

Denote the lower bound as

$$J(\pi_i, s_0; \boldsymbol{\pi}_{-i}) := \sum_{t=0}^{T} \mathbb{E}[r_t^i(s_t, a_t) + H(\pi_i(a_t^i|s_t, a_t^{-i})) \\ - \text{KL}(\rho(a_t^{-i}|s_t)||\boldsymbol{\pi}_{-i}(a_t^{-i}|s_t))], \quad (3)$$

where the expectation is taken with respect to $a_t^i \sim \pi_i(\cdot|s_t, a_t^{-i})$, $a_t^{-i} \sim \rho(\cdot|s_t)$, $s_{t+1} \sim P(\cdot|s_t, a_t^i, a_t^{-i})$. It is known as the evidence lower bound (ELBO) in the context of variational inference [3]. We maximize the ELBO to approximate the best response policy.

We denote the solution to this problem as the joint policy $\boldsymbol{\pi}^*$. And $\boldsymbol{\pi}^*$ satisfies $P(O_{0:T}^i = 1|\pi_i^*, \boldsymbol{\pi}_{-i}^*) \geq P(O_{0:T}^i = 1|\pi_i, \boldsymbol{\pi}_{-i}^*)$ for all $i \in \mathcal{N}$. Then we show the relation between $\boldsymbol{\pi}^*$ and the Nash equilibrium in the following theorem.

**Theorem 4.2.** $\boldsymbol{\pi}^*$ is a $T \log \max_i |A_i|-$Nash equilibrium.

Proof. Denote

$$\tilde{J}(\pi_i, s_0; \boldsymbol{\pi}_{-i}) = \mathbb{E}_{\pi_i}\left[\sum_{t=0}^{T} r^i(s_t, a_t^i, a_t^{-i})\right],$$

where the expectation is taken with respect to $s_{t+1} \sim P(\cdot|s_t, a_t^i, a_t^{-i})$, $a_t^i \sim \pi_i(\cdot|s_t, a_t^{-i})$, $a_t^{-i} \sim \rho(\cdot|s_t)$. Assume that $\tilde{\pi}_i = \arg\max_{\pi_i} \tilde{J}(\pi_i, s_0; \boldsymbol{\pi}_{-i}^*)$ for all $i \in \mathcal{N}$. Then we prove the conclusion by proving the inequality.

$$\tilde{J}(\pi_i^*, s_0; \boldsymbol{\pi}_{-i}) \leq \tilde{J}(\tilde{\pi}_i, s_0; \boldsymbol{\pi}_{-i}) \leq J(\pi_i^*, s_0; \boldsymbol{\pi}_{-i}) \\ \leq \tilde{J}(\pi_i^*, s_0; \boldsymbol{\pi}_{-i}) + T \log \max_{i \in \mathcal{N}} |A_i|$$

The first inequality is from the definition of the best response.

The second inequality is from the non-negativity of entropy.

$$\tilde{J}(\tilde{\pi}_i, s_0; \boldsymbol{\pi}_{-i}) \\ \leq \tilde{J}(\tilde{\pi}_i, s_0; \boldsymbol{\pi}_{-i}) + \mathbb{E}\big[\sum_{t=0}^{T} H(\tilde{\pi}_i(a_t^i|s_t, a_t^{-i}))\big] \\ \leq J(\pi_i^*, s_0; \boldsymbol{\pi}_{-i})$$

The third inequality is from the maximum of entropy: $H(\pi_i(a_t^i|s_t, a_t^{-i})) \leq \log |\mathcal{A}_i|$. Therefore, $\boldsymbol{\pi}^*$ is an $T \log \max_i |A_i|$-Nash equilibrium. □

Theorem 4.2 plays an important role in the framework. It points out that the joint policy derived by the variational inference could approximate Nash equilibrium.

So far the theoretical results are established in the finite-horizon setting, but these results could be generalized to the infinite-horizon setting. In order to ensure the algorithm applicable to the infinite-horizon setting and convergent backup, we modify the transition kernel $\tilde{p}(s_{t+1}|s_t, a_t) = \gamma p(s_{t+1}|s_t, a_t)$, and add an absorbing state $\bar{s}$ such that $\tilde{p}(\bar{s}|s, a) = 1 - \gamma$ for all $s \in \mathcal{S}$ and $a \in \mathcal{A}$ [11].

## 5 Variational Stochastic Game: Realization

In this section, we propose a series of methods for different equilibrium concepts based on the framework provided in Section 4. Since there are only a few differences among these methods, we will give a detailed introduction for inference for Nash equilibrium and only introduce the modifications of other methods. The overview of the differences among these methods is provided in Table 1.

### 5.1 Inference for Nash equilibrium

Based on the framework, we propose a decentralized algorithm named variational policy gradient (VPG) to solve the stochastic game. We analyze the performance difference of policy resulted from the inaccurate opponent modeling, and give a uniform bound applicable to all opponent model methods.

*5.1.1 Variational policy gradient.* We first define the action-value function and value function.

**Definition 5.1.** Given a joint policy $\pi$, the action-value function is defined as follows.

$$Q^i(s_t, a_t^i, a_t^{-i}; \pi) = r_t^i(s_t, a_t) + \log \pi_{-i}(a_t^{-i}|s_t)$$
$$+ \mathbb{E}\left[\sum_{k=t+1}^{T} \gamma^{k-t}(r_k^i(s_k, a_k) + H(\pi_i(a_k^i|s_k, a_k^{-i})))\right.$$
$$\left. - \mathrm{KL}(\rho(a_k^{-i}|s_k)||\pi_{-i}(a_k^{-i}|s_k)))\right]$$

where the expectation is taken with respect to $a_k^i \sim \pi_i(\cdot|s_k, a_k^{-i})$, $a_k^{-i} \sim \rho(\cdot|s_k)$, $s_{k+1} \sim P(\cdot|s_k, a_k^i, a_k^{-i})$. And the value function is

$$V^i(s; \pi) = \mathbb{E}[Q^i(s, a^i, a^{-i}; \pi) - \log \pi_i(a^i|s, a^{-i})\rho(a_t^{-i}|s_t)],$$

where the expectation is taken with respect to $a^i \sim \pi_i(\cdot|s, a^{-i})$, $a^{-i} \sim \rho(\cdot|s)$.

Note that the definition of the action-value function requires that we approximate the policy of opponents $\pi_{-i}(a_t^{-i}|s_t)$ with agent $i$'s opponent model $\rho(a_t^{-i}|s_t)$. Here we don't specify the method to update $\rho(a_t^{-i}|s_t)$. The following analysis applies to any opponent model method.

Using the variational inference, we maximize the ELBO to get the best response of agent $i$.

$$J(\pi_i, s_0; \pi_{-i}) = \mathbb{E}_{s_0 \sim P(s_0)}\left[V^i(s_0; \pi)\right]$$
$$= \mathbb{E}_{s_0 \sim P(s_0)}\left[\mathbb{E}[Q^i(s_0, a^i, a^{-i}; \pi) - \log \pi_i(a^i|s_0, a^{-i})] \right.$$
$$\left. - \log \rho(a^{-i}|s_0)\right]$$
$$= \mathbb{E}_{s_0 \sim P(s_0)}\left[Z + H(\rho(\cdot|s_0)) \right.$$
$$\left. - \mathbb{E}\left[\mathrm{KL}\left(\pi_i(a^i|s_0, a^{-i})\middle\|\frac{\exp(Q^i(s_0, a^i, a^{-i}; \pi))}{Z}\right)\right]\right],$$

where $Z = \sum_{a^i \in A_i} \exp(Q^i(s_0, a^i, a^{-i}; \pi))$. Since the KL divergence is non-negative, we have the following proposition.

**Proposition 5.2.** *The best response policy is in the form of*

$$\pi_i^*(a^i|s, a^{-i}) = \frac{\exp(Q^i(s, a^i, a^{-i}; \pi))}{\sum_{a^i \in A_i} \exp(Q^i(s, a^i, a^{-i}; \pi))}$$

Proposition 5.2 shows that the best response policy is in the form of $\pi_i^\theta(a|s, a^{-i}) = \mathrm{softmax}(\theta_{i,s,a,a^{-i}})$, i.e. the best response policy is softmax policy parameterized [2]. We use the natural policy gradient (NPG) method [9] to derive the best response policy.

**Proposition 5.3.** *Denote $\theta^{(t)}$ the $t$-th iterate and $\pi^{(t)} = \mathrm{softmax}(\theta_{s,a})$. For each agent $i$, state $s$, and action $a$, the NPG update rule can be written as*

$$\pi_i^{(t+1)}(a \mid s, a^{-i}) = \frac{1}{Z^{(t)}(s)} \left(\pi^{i,(t)}(a \mid s, a^{-i})\right)^{1-\frac{\eta}{1-\gamma}} \exp\left(\frac{\eta Q^{i,(t)}(s, a, a^{-i}; \pi^{(t)})}{1-\gamma}\right). \quad (4)$$

*where $\eta$ is the learning rate.*

PROOF. We first define the discounted state visitation distribution $d^\pi(s) = (1-\gamma)\sum_{t=0}^{\infty}\gamma^t P(s_t = s)$. Denote $\theta_i^{(t)}$ is the parameter vector of policy $\pi_i^{(t)}$. And the element of $\theta^{(t)}$ is the approximation of action-value function, i.e. $\theta_{i,s,a,a^{-i}}^{(t)} = Q^i(s, a, a^{-i}; \pi^{(t)})$. The update rule of $\theta$ is

$$\theta_{i,s,a,a^{-i}}^{(t+1)} = \theta_{i,s,a,a^{-i}}^{(t)} + \eta \left[\mathcal{F}(\theta_i^{(t)})^\dagger V^i(s; \pi^{(t)})\right]$$
$$= \theta_{i,s,a,a^{-i}}^{(t)} + \frac{\eta}{1-\gamma}\left(Q^i(s, a, a^{-i}; \pi^{(t)}) \right.$$
$$\left. - \log \pi(a|s, a^{-i})\rho(a^{-i}|s) - V^i(s; \pi^{(t)})\right),$$

where $\mathcal{F}(\theta_i^{(t)})^\dagger$ is the pseudo-inverse of the Fischer information matrix

$$\mathcal{F}(\theta_i^{(t)}) = \mathbb{E}_{s \sim d^{\pi^{(t)}}(\cdot), a^{-i} \sim \rho(\cdot|s), a^i \sim \pi_i^{(t)}(\cdot|a^{-i},s)}\left[\nabla_{\theta_i^{(t)}} \log \pi_i^{(t)}(a^i|s, a^{-i}) \log \pi_i^{(t)}(a^i|s, a^{-i})^T\right].$$

Table 1: The modification of framework to different games.

|  | Opponent Model | Policy |
|---|---|---|
| Nash equilibrium in general-sum stochastic game | $\rho(a^{-i}\|s)$ | $\pi(a^i\|s, a^{-i})$ |
| Nash equilibrium in mean field game | $\mathcal{L}(s, a)$ | $\pi(a\|s, \mathcal{L})$ |
| Correlated equilibrium in general-sum stochastic game | $\rho(a^{-i}\|s)$ | $\pi(a_t^i\|\tilde{s}_t, a_t^{-i}) = \sum_{\omega_t \in \Omega} \pi(a_t^i\|\tilde{s}_t, a_t^{-i}, \omega_t)\sigma(\omega_t)$ |
| Nash equilibrium in zero sum stochastic game | None | $\pi(a^i\|s, a^{-i})$ |

**Algorithm 1** Variational Policy Gradient (VPG)
─────────────────────────────────────────────
**Require:** Learning rate $\eta$
　Initialise opponent model $\rho$.
　Initialise policy $\pi^{i,(0)}$ for all agent $i \in \mathcal{N}$.
　Initialise the replay buffer $M$.
　**for** $k = 1, 2, \ldots$ **do**
　　**for** Each agent $i \in \mathcal{N}$ **do**
　　　For the current state $s_t$, $a_t^i \sim \pi^i(\cdot|s_t) = \sum_{a_t^{-i}} \rho(a_t^{-i}|s_t)\pi^i(\cdot|s_t, a_t^i, a_t^{-i})$.
　　　Observe next state $s_{t+1}$, opponent action $a_t^{-i}$ and reward $r_t^i$ and save the experience in the reply buffer.
　　　Update opponent model.
　　**end for**
　　**for** Each agent $i \in \mathcal{N}$ **do**
　　　Compute the best response policy using Equation (4).
　　**end for**
　**end for**
─────────────────────────────────────────────

Hence the update rule of policy is

$$\pi_i^{(t+1)}(a|s, a^{-i}) \propto \exp(\theta_{i,s,a,a^{-i}}^{(t+1)})$$
$$= \exp\left(\theta_{i,s,a,a^{-i}}^{(t)} + \frac{\eta}{1-\gamma}\left(Q^i(s, a, a^{-i}; \boldsymbol{\pi}^{(t)})\right.\right.$$
$$\left.\left. - \log \pi(a|s, a^{-i})\rho(a^{-i}|s) - V^i(s; \boldsymbol{\pi}^{(t)})\right)\right)$$
$$\propto \left(\pi_i^{(t)}(a|s, a^{-i})\right)^{1-\frac{\eta}{1-\gamma}} \exp\left(\frac{\eta}{1-\gamma}Q^i\left(s, a, a^{-i}; \boldsymbol{\pi}^{(t)}\right)\right).$$
□

Then we propose the variational policy gradient (VPG) algorithm. The pseudo-code of VPG is listed in the Algorithm 1.

Then we will prove that VPG converges to Nash equilibrium in the Markov potential game.

**Definition 5.4.** *Markov Potential Game (MPG) is a Markov decision process that there exists a function $\Phi(s; \pi_i, \boldsymbol{\pi}_{-i}) : \Pi \to \mathbb{R}$, with $s \in \mathcal{S}$, so that*
$$\tilde{V}^i(s; \pi_i, \boldsymbol{\pi}_{-i}) - \tilde{V}^i(s; \pi_i', \boldsymbol{\pi}_{-i})$$
$$= \Phi(s; \pi_i, \boldsymbol{\pi}_{-i}) - \Phi(s; \pi_i', \boldsymbol{\pi}_{-i}),$$
*for all agents $i \in \mathcal{N}$, states $s \in \mathcal{S}$ and policies $\pi_i, \pi_i' \in \Pi_i, \boldsymbol{\pi}_{-i} \in \Pi_{-i}$. Here $\tilde{V}^i(s; \pi_i, \boldsymbol{\pi}_{-i})$ is value function with accurate opponent modeling.*

The first step is to prove that the estimation error of the opponent is bounded. Modeling the opponent will result in an estimation error of the action-value function. The following proposition gives the upper bound of estimation error.

**Proposition 5.5.** *Suppose that $\text{KL}(\rho(\cdot|s)||\boldsymbol{\pi}_{-i}(\cdot|s)) < \epsilon_\rho$ for all $s \in \mathcal{S}$. Without loss of generality, the reward function $|r^i(s, \boldsymbol{a})| \le 1$, $\forall s \in \mathcal{S}, \boldsymbol{a} \in \mathcal{A}, i \in \mathcal{N}$. Denote the action-value function derived using the opponent model as $\hat{Q}^i(s, \boldsymbol{a}; \boldsymbol{\pi})$. Then we have that $\max_{s \in \mathcal{S}, \boldsymbol{a} \in \mathcal{A}, i \in \mathcal{N}} |Q^i(s, \boldsymbol{a}; \boldsymbol{\pi}) - \hat{Q}^i(s, \boldsymbol{a}; \boldsymbol{\pi})| \le \delta$ where*

$$\delta := \frac{2(1 + \log |\mathcal{A}_i|)}{(1-\gamma)^2}\sqrt{\frac{1}{2}\epsilon_\rho} + \frac{\epsilon_\rho}{1-\gamma}$$

.

The proof is deferred to Appendix A.1.

The second step is to derive the convergence of VPG with exact opponent modeling. We first show the equivalence between VPG and the global NPG on the potential function. Then we will prove the convergence of VPG using the smoothness of the potential function.

Note that the gradient of the value functions equals the potential function and agents update their policy independently. Hence VPG is equivalent to running Natural Policy Gradient (NPG) on the potential function, which is shown in the following proposition.

**Proposition 5.6.** *Consider the global NPG dynamic on the potential function: $\theta_s^{(t+1)} = \theta_s^{(t)} + \eta \mathcal{F}^\dagger(\theta_s^{(t)})\nabla_{\theta_s}\Phi \; \forall s \in \mathcal{S}$, where $\mathcal{F}^\dagger(\theta_s) = \mathbb{E}[\nabla_{\theta_s}\log \boldsymbol{\pi}^{\theta_s}(\boldsymbol{a}|s)\nabla_{\theta_s}\log \boldsymbol{\pi}^{\theta_s}(\boldsymbol{a}|s)^T]^\dagger$ is the pseudo-inverse of the Fischer information matrix. $\boldsymbol{\pi}^{\theta_s}(\boldsymbol{a}|s) = \prod_{i \in \mathcal{N}} \mathbb{E}_{a^{-i} \sim \rho(\cdot|s)}[\pi_i(a^i|s, a^{-i})]$. VPG has the same dynamics as global NPG.*

The proof is deferred to Appendix A.2. After showing the connection of VPG and the NPG on the potential function, we next show the smoothness of the potential function in the following lemma.

**Lemma 5.7.** *The potential function $\Phi$ is L-smooth with the constant $L = \frac{2(n+1)^2}{(1-\gamma)^3} + 2(n^2 + n + 1)\frac{1+\log \max_{i \in \mathcal{N}}|A_i|}{(1-\gamma)^2} + \frac{3n+2}{1-\gamma}$.*

The proof is deferred to Appendix A.3. Using Lemma 5.7, the potential function $\Phi(s; \boldsymbol{\pi}^{(t)})$ is non-decreasing if the learning rate is $\frac{1}{L}$ [4]. We finally give the convergence of VPG.

**Theorem 5.8.** *VPG converges to a fixed point, which is $\epsilon$-Nash equilibrium of MPG, where $\epsilon = \delta + \frac{\log |A|}{1-\gamma}$.*

Theorem 5.8 ensures the applicability of VPG for solving MPG. VPG does not involve a certain opponent modeling method. The next question is how to model the opponent using variational inference.

*5.1.2 Opponent modeling.* Since the agent has no knowledge about the optimality probabilities of other agents, we derive a variational inference method to model opponents. To model the behavior of agent $j$, we factorize the auxiliary distribution over states and actions $q(a^j_{0:\infty}, a^{-j}_{0:\infty}, s_{0:\infty})$ in a following way.

$$q(a^j_{0:\infty}, a^{-j}_{0:\infty}, s_{0:\infty})$$
$$= P(s_0) \prod_{t=0}^{\infty} q(a^j_t|s_t) q(a^{-j}_t|s_t) P(s_{t+1}|s_t, a^j_t, a^{-j}_t)$$
$$= P(s_0) \prod_{t=0}^{\infty} \rho_j(a^j_t|s_t) \rho_{-j}(a^{-j}_t|s_t) P(s_{t+1}|s_t, a^j_t, a^{-j}_t)$$

Denote $Q^j_\rho(s_t, \boldsymbol{a}_t; \rho)$ as the soft action-value function of agent $j$.

$$Q^j_\rho(s_t, \boldsymbol{a}_t; \rho) = r^j(s_t, \boldsymbol{a}_t) - \text{KL}(\rho_{-j}(a^{-j}_t|s_t) || \boldsymbol{\pi}_{-j}(a^{-j}_t|s_t)$$
$$+ \mathbb{E}[\sum_{i=t+1}^{\infty} \gamma^{i-t}(r^j_i(s_i, \boldsymbol{a}_i) - \text{KL}(\rho_j(a^j_i|s_i) || \pi_j(a^j_i|s_i)))],$$

where the expectation is taken with respect to $\boldsymbol{a}_i \sim q(\cdot|s_i))$, $s_i \sim P(\cdot|s_{i-1}, \boldsymbol{a}_{i-1})$. Then we can derive the optimal opponent model for agent $j$.

**Proposition 5.9.** *The optimal opponent model for agent $j$ is*

$$\rho^*_j(a^j|s) = \frac{\hat{\pi}_j(a^j|s) \exp(\mathbb{E}_{a^{-j} \sim \rho_{-j}}[Q^j_\rho(s, \boldsymbol{a}; \rho)])}{\mathbb{E}_{a^j \sim \hat{\pi}_j(\cdot|s)}\left[\exp(\mathbb{E}_{a^{-j} \sim \rho_{-j}}[Q^j_\rho(s, \boldsymbol{a}; \rho)])\right]} \quad (5)$$

*where $\hat{\pi}_j(a^j|s)$ is the prior of policy $\pi_j(a^j|s)$.*

The proof is deferred to Appendix A.5.

Since the agent $i$ does not know the reward of the agent $j$, we have to find a function $\hat{r}^j$ to estimate $r^j$. The objective of optimizing $\hat{r}^j$ is to minimize the KL divergence between the optimal opponent model derived by estimated reward function $\hat{r}^j$ and the history data of agent $j$. Let $\tau_j = \{s_0, a^j_0, a^{-j}_0, s_1, a^j_1, a^{-j}_1, \cdots\}$ be the history data of agent $j$. The probability of generating $\tau_j$ through taking actions $\{a^j_0, a^j_1, \cdots\}$ is

$$P(\tau_j|a^j_{0:\infty}) = P(s_0) \prod_{t=1}^{\infty} P(s_t|s_{t-1}, a^j_{t-1}, a^{-j}_{t-1}) \rho_{-j}(a^{-j}_{t-1}|s_{t-1}).$$

And the probability of generating $\tau_j$ by the opponent model is

$$\rho_j(\tau_j) = P(s_0) \prod_{t=1}^{\infty} P(s_t|s_{t-1}, a^j_{t-1}, a^{-j}_{t-1}) \rho^*_j(a^j_{t-1}|s_{t-1})$$
$$\rho_{-j}(a^{-j}_{t-1}|s_{t-1}).$$

Then the objective to optimize $\hat{r}^j$ is

$$\text{KL}(P(\tau_j) || \rho_j(\tau_j)) = \mathbb{E}\left[\sum_{t=0}^{\infty} -\gamma^t \hat{r}^j(s_t, a^j_t, a^{-j}_t)\right]$$
$$+ \log \mathbb{E}_{a^j \sim \rho_j}\left[\mathbb{E}_{a^{-j} \sim \rho_{-j}}[\exp(Q^j_\rho(s, \boldsymbol{a}; \rho))]\right],$$

where the first expectation is taken with respect to state $s_t \sim P(s_t|s_{t-1}, a^j_{t-1}, a^{-j}_{t-1})$. It is difficult to calculate the optimal opponent model because $\mathbb{E}_{a^j \sim \rho_j}\left[\exp(Q^j_\rho(s, \boldsymbol{a}; \rho))\right]$ is difficult to estimate. In addition, we use a sample-based method for estimating

$\mathbb{E}_{a^j \sim \rho_j}\left[\exp(Q^j_\rho(s, \boldsymbol{a}; \rho))\right]$.

$$\text{KL}(P(\tau_j)||\rho_j(\tau_j))$$
$$= \mathbb{E}\left[\sum_{t=0}^{\infty} -\gamma^t \hat{r}^j(s_t, a^j_t, a^{-j}_t)\right] \quad (6)$$
$$+ \log \mathbb{E}_{\tau_j \sim \rho(\tau_j)}\left[\frac{\exp(\sum_{t=0}^{\infty} \gamma^t \hat{r}^j(s_t, \boldsymbol{a}_t))}{\rho(\tau_j)}\right]$$

If $\hat{r}^j_\psi$ is parameterized by $\psi$, the gradient of $\text{KL}(P(\tau_j)||\rho_j(\tau_j))$ with respect to $\psi$ is

$$\frac{d\text{KL}(P(\tau_j)||\rho_j(\tau_j))}{d\psi} = \mathbb{E}\left[\sum_{t=0}^{\infty} -\gamma^t \frac{\hat{r}^j_\psi(s_t, a^j_t, a^{-j}_t)}{d\psi}\right] \quad (7)$$
$$+ \frac{1}{Z} \mathbb{E}_{\tau_j \sim \rho(\tau_j)}\left[w_j \frac{d\sum_{t=0}^{\infty} \gamma^t \hat{r}^j_\psi(s_t, \boldsymbol{a}_t)}{d\psi}\right],$$

where $w_j = \frac{\exp(\sum_{t=0}^{\infty} \gamma^t \hat{r}^j_\psi(s_t, \boldsymbol{a}_t))}{\rho(\tau_j)}$ and $Z = \mathbb{E}_{\tau_j \sim \rho(\tau_j)}[w_j]$.

VPG is for tabular cases and is impractical in problems with high dimensions or continuous action. To handle the problems, we propose the variational actor-critic method, which can be implemented in a complex continuous environment. We use neural-network to parameterize the policy $\pi^\theta$, opponent model $\rho^\phi$, the action-value function $Q_\omega$, and the reward function $r_\psi$.

The objective to optimize the policy $\pi^\theta$ is to minimize the KL divergence

$$J_\pi(\theta; s) = \mathbb{E}_{a^{-i} \sim \rho(\cdot|s)}$$
$$\left[\text{KL}\left(\pi^\theta_i(\cdot|s) || \exp(Q^i_\omega(s, \cdot, a^{-i}) - V^i(s))\right)\right]. \quad (8)$$

The objective to optimize the action-value function $Q_\omega$ is to minimize:

$$J_Q(\omega) = \mathbb{E}_{(s_t, a^i_t, a^{-i}_t) \sim \mathcal{D}}\left[\frac{1}{2}\left(Q^i_\omega\left(s_t, a^i_t, a^{-i}_t\right)\right.\right.$$
$$\left.\left. - r^i\left(s_t, a^i_t, a^{-i}_t\right) - \gamma \mathbb{E}_{s_{t+1} \sim p_s}\left[\bar{V}(s_{t+1})\right]\right)^2\right], \quad (9)$$

with

$$\bar{V}^i(s_{t+1}) = Q^i_{\bar{\omega}}\left(s_{t+1}, a^i_{t+1}, \hat{a}^{-i}_{t+1}\right) - \log \rho_\phi\left(\hat{a}^{-i}_{t+1} | s_{t+1}\right)$$
$$- \log \pi_\theta\left(a^i_{t+1} | s_{t+1}, \hat{a}^{-i}_{t+1}\right) + \log P\left(\hat{a}^{-i}_{t+1} | s_{t+1}\right),$$

where $Q^i_{\bar{\omega}}$ is target Q function.

The gradient of (8) with respect to $\theta$ is

$$\nabla_\theta J_\pi(\theta; s) = \mathbb{E}_{a^{-i} \sim \rho(\cdot|s)}[\nabla_\theta \log \pi^\theta_i(a|s)$$
$$+ (\nabla_a \pi^\theta_i(a|a^{-j}, s) - \nabla_a Q^i(s, a, a^{-i}))\nabla_\theta f_\theta(\epsilon; s, a^{-i})] \quad (10)$$

where $a$ is evaluated at $f_\theta(\epsilon; s, a^{-i})$. The gradient of (9) with respect to $\omega$ is

$$\nabla_\omega J_Q(\omega) = \nabla_\omega Q^i_\omega\left(s_t, a^i_t, a^{-i}_t\right)\left(Q^i_\omega\left(s_t, a^i_t, a^{-i}_t\right)\right.$$
$$\left. - r^i\left(s_t, a^i_t, a^{-i}_t\right) - \gamma \mathbb{E}_{s_{t+1} \sim p_s}\left[\bar{V}(s_{t+1})\right]\right)$$

Then the pseudo-code of the variational inference actor-critic method named Multi-agent Inference (MAI) is listed in the Algorithm 2.

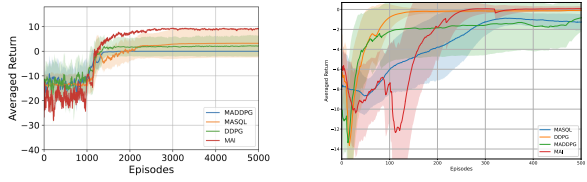

(a) The learning curves of MAI and other baselines in differential game.

(b) The learning curves of MAI and other baselines in non-atomic routing game.

Figure 2: Learning curves in differential game and non-atomic routing game.

**Experiments:** As MAI incorporates entropy regularization naturally, it enjoys stronger exploration ability. We test its exploration ability on a challenging differential game. Differential game is adopted from [22]. The two agents in this game have continuous action space. All the agents share the same reward function depending on the joint action $(a_1, a_2)$ following the equations: $r^1(a^1, a^2) = r^2(a^1, a^2) = \max(f_1, f_2)$, where $f_1 = 0.8 \times \left[ -\left(\frac{a^1+5}{3}\right)^2 - \left(\frac{a^2+5}{3}\right)^2 \right]$, $f_2 = 1.0 \times \left[ -\left(\frac{a^1-5}{1}\right)^2 - \left(\frac{a^2-5}{1}\right)^2 \right] + 10$. This task is highly challenging for most continuous gradient-based reinforcement learning algorithms, as the gradient updates often guide the agent towards suboptimal solutions. The training process includes 200 episodes with 25 steps per episode. We compare MAI with MADDPG [14], MASQL [26] and independent DDPG [13] in this task. The learning curves are shown in Figure 2a. Only MAI shows the capability of converging to the global optimum, while most of the baselines can only reach the sub-optimal point. The result illustrates that the algorithm derived from the variational inference framework has stronger exploration ability.

To evaluate that MAI can converge on the Markov potential game, we conduct MAI on a Markov potential game task named non-atomic routing game. We adopt the game from [15]. Agents in this game are self-interested. Each agent learns how to split their commodity in a way that maximizes rewards. We compare MAI with MADDPG, MASQL and independent DDPG in this task. The learning curves are shown in Figure 2b. MAI achieves a higher return than other baselines. The learning curve of MAI is smoother and other algorithms suffer from high variance.

The result of MASQL can be viewed as an ablation study of MAI. As MASQL does not use the opponent modeling, it is not able to converge to the global optimum. It is notable that the application of MAI is not limited to the cases with continuous action space. We also test MAI on games where agents share the same reward function. The opponent modeling can be applied to the cases where agents have different reward functions.

---

**Algorithm 2** Multi-agent Inference (MAI)

Initializing replay buffer $\mathcal{D}$, $\theta$, $\omega$, $\psi$ and $\phi$.
**for** Each episode $d = 1, 2, \cdots$ **do**
  **for** $i \in \mathcal{N}$ **do**
    For current state $s_t$ compute $a_t^{-i} \sim \rho(\cdot|s_t)$, $a_t^i \sim \pi_i(\cdot|s_t, a_t^{-i})$
    Observe next state $s_{t+1}$, opponent action $a_t^{-i}$ and save the new experience in the reply buffer $\mathcal{D}$.
    Update opponent model using Algorithm 3.
    Update $\pi_i$ using (10).
  **end for**
**end for**
**Output:** policy $\pi_i$, $i \in \mathcal{N}$, opponent model $\rho$

---

**Algorithm 3** Opponent modeling (OM)

**Require:** Initial parameters of the reward function $\psi$, trajectory replay buffer $\mathcal{D}$
1: **for** $i = 1, 2, \ldots$ **do**
2:   Sample trajectory $\tau$ from $\mathcal{D}$
3:   **for** $j = 1$ to $N$ **do**
4:     Update $\hat{r}^j(s_t, \boldsymbol{a}_t)$ using Equation (7)
5:     Update $\rho_j(\tau_j)$ using Equation (5)
6:   **end for**
7: **end for**
**Ensure:** Optimized opponent model $\rho$

### 5.2 Inference for mean field Nash equilibrium

Although this algorithm is distributed, Solving $N$-player game is still intractable when the $N$ is large. If all the agents are homogeneous and interchangeable, we can alleviate this problem. The states and actions of other agents can be reduced into a joint distribution of the population state-action pair $\mathcal{L}_t = \mathbb{P}_{s_t, a_t} \in \mathcal{P}(\mathcal{S} \times \mathcal{A})$, which is named mean field. The mean field follows the Kolmogorov equation

$$\mathcal{L}_t(s, a) = q_i^k(a|s) \sum_{s_t \in \mathcal{S}} \sum_{a_t \in \mathcal{A}} P(s|s_{t-1}, a_{t-1}, \mathcal{L}_{t-1}) \mathcal{L}_{t-1}(s_t, a_t).$$

Given the mean field, the objective to optimize the policy $\pi$ is to maximize the likelihood $P(\mathcal{O}_{0:T} = 1|\pi, \mathcal{L}_{0:T})$. Denote the trajectory $\tau = \{s_0, a_0, \mathcal{L}_0, s_1, a_1, \mathcal{L}_1, \cdots\}$. The $q(\tau)$ is the probability to generate the trajectory $\tau = \{s_0, a_0, \mathcal{L}_0, s_1, a_1, \mathcal{L}_1, \cdots\}$ by the policy $\pi$. The $q(\tau)$ can be factorized as follows.

$$q(\tau) = P(s_0) \prod_{t=0}^{T-1} q(a_t|s_t) P(s_{t+1}|s_t, a_t, \mathcal{L}_t)$$

$$= P(s_0) \prod_{t=0}^{T-1} \pi(a_t|s_t) P(s_{t+1}|s_t, a_t, \mathcal{L}_t)$$

To stabilize the learning process, we introduce a prior distribution policy that uses information from past iterations.

$$\log P(O_{0:T} = 1|\pi, \mathcal{L}_{0:T})$$
$$\geq \sum_{s_{0:T} \in \mathcal{S}, a_{0:T} \in \mathcal{A}} q(\tau) \log \frac{P(O_{0:T} = 1|\pi, \mathcal{L}_{0:T})}{q(\tau)}$$
$$= \sum_{t=0}^{T} \mathbb{E}\left[r(s_t, a_t, \mathcal{L}_t) - \log \frac{\pi_t(a_t|s_t)}{\hat{\pi}_t(a_t|s_t)}\right]$$

where $\hat{\pi}_t(a_t|s_t)$ is the prior policy. The objective of agents $i$ is to maximize $\mathbb{E} \sum_{t=0}^{T}\left[r(s_t^i, a_t^i, \mathcal{L}_t) - \log \pi_i(a_t|s_t, \mathcal{L}_t) + \log \hat{\pi}(a_t|s_t, \mathcal{L}_t)\right]$, where the expectation is taken with respect to $s_{t+1} \sim P(\cdot|s_t, a_t, \mathcal{L}_t)$, $a_t \sim \hat{\pi}(\cdot|s_t, \mathcal{L}_t)$. Denote the action-value function

$$Q_t(s_t, a_t, \mathcal{L}_t) = r(s_t, a_t, \mathcal{L}_t)$$
$$+ \sum_{s \in \mathcal{S}} P(s|s_t^i, a_t^i, \mathcal{L}_t) \mathbb{E}_{a \sim \pi(\cdot|s, \mathcal{L}_{t+1})}\left[Q_{t+1}(s, a, \mathcal{L}_{t+1})\right.$$
$$\left. - \log \pi(a_t|s_t, \mathcal{L}_t) + \log \hat{\pi}(a_t|s_t, \mathcal{L}_t)\right] \quad (11)$$

with the terminal condition $Q_T(s_T, a_T, \mathcal{L}_T) = r(s_T, a_T, \mathcal{L}_T)$. The optimal policy has the closed form

$$\pi_t(a|s) = \frac{\hat{\pi}_t(a|s) \exp(Q_t(s, a, \mathcal{L}))}{\sum_{a \in \mathcal{A}} \hat{\pi}_t(a|s) \exp(Q_t(s, a, \mathcal{L}))} \quad (12)$$

---

**Algorithm 4** Mean field Bayesian Q-learning

**Require:** Initial $\mathcal{L}^0$ and initial prior policy $\hat{\pi}$.
   **for** k=1, 2, $\cdots$ **do**
      Compute the soft Q function $Q_t(s, a, \mathcal{L}_t^k)$ using (11).
      Compute $\pi_t^k$ using (12).
      Compute mean field $\mathcal{L}_t(s, a)$ induced by $\pi_t^k$ from the simulator.
   **end for**

---

Then we propose our Algorithm 4 named Mean field Bayesian Q-learning. It is notable that mean field Bayesian Q-learning is consistent with RelEnt iteration [5]. Here we theoretically derive the relative entropy regularizer directly rather than treat it as a heuristic modification.

## 5.3 Inference for correlated equilibrium

In the correlated equilibrium, we assume that all agents take actions according to a publicly observed random variable, namely the correlated signal $\omega$. The graphical model for solving correlated equilibrium is shown in Figure 3. We assume that the correlated signal is sampled from a distribution $\sigma(\omega)$ over the signal space $\Omega$.

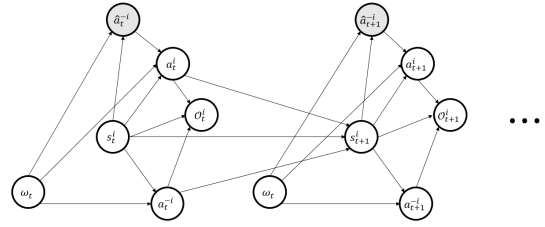

**Figure 3: The graphic model of correlated equilibrium. Compared with Figure 1, the correlated signal $\omega$ is added. We assume that all the agents choose actions dependent on this publicly observed random variable $\omega$.**

The trajectory probability can be factorized in a such way.

$$q(a_{0:T}^i, a_{0:T}^{-i}, s_{0:T})$$
$$= P(s_0) \prod_{t=0}^{T} \sum_{\omega_t \in \Omega} \sigma(\omega_t) P(s_{t+1}|s_t, \boldsymbol{a}_t) q(a_t^i|a_t^{-i}, s_t, \omega_t)$$
$$q(a_t^{-i}|s_t, z_i)$$
$$= P(s_0) \prod_{t=0}^{T} \sum_{\omega_t \in \Omega} \sigma(\omega_t) P(s_{t+1}|s_t, \boldsymbol{a}_t) \pi(a_t^i|s_t, a_t^{-i}, \omega_t)$$
$$\rho(a_t^{-i}|s_t, \omega_t)$$

The correlated equilibrium can be solved using the methods for solving Nash equilibrium. The correlated signal $\omega$ can be augmented into the state $s$. And the transition dynamic of the augmented state $\tilde{s} = (s, \omega)$ is $P(\tilde{s}_{t+1}|\tilde{s}_t, \boldsymbol{a}_t) = \sigma(\omega_{t+1}) \sum_{\omega_t \in \Omega} P(s_{t+1}|s_t, \boldsymbol{a}_t) \sigma(\omega_t)$.

## 5.4 Inference for Nash equilibrium in zero-sum game

There are two player in the zero-sum game and their rewards are opposite. Given the reward function $r^i(s, a^i, a^{-i})$ of agent $i$, the reward of agent $-i$ is $r^{-i}(s, a^i, a^{-i}) = r^i(s, a^i, a^{-i})$. Therefore, $\log P(O^i|s, a^i, a^{-i}) = -\log P(O^{-i}|s, a^i, a^{-i})$. The objective of agent $i$ is the same as the stochastic game, while the objective of the other agent is also known for agent $i$. Hence it is unnecessary for agent $i$ to estimate the reward function of agent $-i$. Under the setting of zero-sum game, the optimal policy of agent $-i$ is

$$\rho_{-i}^*(a^{-i}|s) = \frac{\hat{\pi}_{-i}(a^{-i}|s) \exp(Q_\rho^{-i}(s, \boldsymbol{a}; \rho))}{\mathbb{E}_{a^{-i} \sim \rho^{-i}}\left[\exp(Q_\rho^{-i}(s, \boldsymbol{a}; \rho))\right]}$$
$$= \frac{\hat{\pi}_{-i}(a^{-i}|s) \exp(-Q^i(s, \boldsymbol{a}; \rho))}{\mathbb{E}_{a^{-i} \sim \rho^{-i}}\left[\exp(-Q^i(s, \boldsymbol{a}; \rho))\right]}$$

The opponent policy is updated after the action-value function of agent $i$ is updated, which simplifies the process of opponent modeling.

## 6 Conclusion

In this paper, we propose a unified variational inference framework for solving general-sum stochastic games, which builds the connection between game theory and probability inference. We prove that the optimal policy under our framework is an $\epsilon$-Nash equilibrium. Leveraging this unified framework, we instantiate different methods to solve Nash equilibrium in the general-sum stochastic game, Nash equilibrium in the mean-field game, correlated equilibrium in the general-sum stochastic game, and Nash equilibrium in the zero-sum stochastic game. We prove that our method can converge in the Markov potential game. We also propose an algorithm to model opponents using variational inference. Furthermore, the proposed opponent modeling algorithm enables agents to reason about the behaviors of others, even in competitive environments with unaligned objectives.

## Acknowledgments

Haifeng Zhang thanks the support of National Science and Technology Major Project 2022ZD0116404.

## A  Proof

### A.1  Proof of Proposition 5.5

Proof. From Pinsker's inequality,

$$D_{TV}(\rho(\cdot|s), \pi^{-i}(\cdot|s)) \leq \sqrt{\frac{1}{2}\mathrm{KL}(\rho(\cdot|s)||\pi^{-i}(\cdot|s))} \leq \sqrt{\frac{1}{2}\epsilon_\rho}.$$

Denote the value function derived using the opponent model as $\hat{V}^i(s; \boldsymbol{\pi})$. Define $P_\rho^{\boldsymbol{\pi}}(s'|s) := \mathbb{E}_{a^{-i} \sim \rho(\cdot|s)}[\sum_{a \in \mathcal{A}^i} P(s'|s, a, a^{-i}) \pi(a|s, a^{-i})]$.

$$|V^i(s; \boldsymbol{\pi}) - \hat{V}^i(s; \boldsymbol{\pi})|$$
$$\leq 2(1 + \log |\mathcal{A}_i|) D_{TV}(\rho(\cdot|s), \pi^{-i}(\cdot|s))$$
$$+ \gamma |\mathbb{E}_{s' \sim P_\rho^{\boldsymbol{\pi}}(s'|s)} V^i(s'; \boldsymbol{\pi}) - \mathbb{E}_{s' \sim P_{\boldsymbol{\pi}^{-i}}^{\boldsymbol{\pi}}(s'|s)} V^i(s'; \boldsymbol{\pi})|$$
$$+ \gamma |\mathbb{E}_{s' \sim P_\rho^{\boldsymbol{\pi}}(s'|s)} [V^i(s'; \boldsymbol{\pi}) - \hat{V}^i(s'; \boldsymbol{\pi})]| - \mathrm{KL}(\rho(\cdot|s) \| \pi^{-i}(\cdot|s))$$
$$\leq 2(1 + \log |\mathcal{A}_i|) D_{TV}(\rho(\cdot|s), \pi^{-i}(\cdot|s))$$
$$+ 2\gamma (\max_{s' \in \mathcal{S}} V^i(s'; \boldsymbol{\pi})) D_{TV}(P_\rho^{\boldsymbol{\pi}}(s'|s), P_{\boldsymbol{\pi}^{-i}}^{\boldsymbol{\pi}}(s'|s))$$
$$+ \gamma \max_{s' \sim \mathcal{S}} |V^i(s'; \boldsymbol{\pi}) - \hat{V}^i(s'; \boldsymbol{\pi})| - \mathrm{KL}(\rho(\cdot|s) \| \pi^{-i}(\cdot|s))$$
$$\leq 2(1 + \log |\mathcal{A}_i| + \frac{\gamma(1 + \log |\mathcal{A}_i|)}{1 - \gamma}) \sqrt{\frac{1}{2} \epsilon_\rho}$$
$$+ \gamma \max_{s' \sim \mathcal{S}} |V^i(s'; \boldsymbol{\pi}) - \hat{V}^i(s'; \boldsymbol{\pi})| + \epsilon_\rho$$
$$= \frac{2(1 + \log |\mathcal{A}_i|)}{1 - \gamma} \sqrt{\frac{1}{2} \epsilon_\rho} + \gamma \max_{s' \sim \mathcal{S}} |V^i(s'; \boldsymbol{\pi}) - \hat{V}^i(s'; \boldsymbol{\pi})| + \epsilon_\rho$$

Then the estimated error of value function can be derived

$$\max_{s \sim \mathcal{S}} |V^i(s; \boldsymbol{\pi}) - \hat{V}^i(s; \boldsymbol{\pi})| \leq \frac{2(1 + \log |\mathcal{A}_i|)}{(1 - \gamma)^2} \sqrt{\frac{1}{2} \epsilon_\rho} + \frac{\epsilon_\rho}{1 - \gamma}$$

Using soft Bellman equation, we have that

$$\max_{s \in \mathcal{S}, \boldsymbol{a} \in \mathcal{A}} |Q^i(s, \boldsymbol{a}; \boldsymbol{\pi}) - \hat{Q}^i(s, \boldsymbol{a}; \boldsymbol{\pi})| \leq \delta$$

□

## A.2 Proof of Proposition 5.6

Proof. We denote $\pi_i(a|s) = \sum_{a^{-i} \in \mathcal{A}^{-i}} \pi_i^{\theta_s}(a|a^{-i}, s) \rho(a^{-i}|s)$. Then the joint policy $\boldsymbol{\pi}^{\theta_s}(\boldsymbol{a}|s) = \prod_{i \in \mathcal{N}} \pi_i^{\theta_s}(a^i|s)$. For all $i, j \in \mathcal{N}$, $i \neq j$.

$$\mathbb{E}_{\boldsymbol{a} \sim \boldsymbol{\pi}^{\theta_s}(\cdot|s)} \left[ \nabla_{\theta_s} \pi_i^{\theta_s}(a^i|s) \nabla_{\theta_s} \pi_j^{\theta_s}(a^j|s)^T \right]$$
$$= \mathbb{E}_{\boldsymbol{a} \sim \boldsymbol{\pi}^{\theta_s}(\cdot|s)} \left[ \nabla_{\theta_s} \pi_i^{\theta_s}(a^i|s) \right] \left[ \nabla_{\theta_s} \pi_j^{\theta_s}(a^j|s)^T \right] = 0$$

Then the Fisher matrix

$$\mathcal{F}(\theta_s) = \mathbb{E}[\nabla_{\theta_s} \log \boldsymbol{\pi}^\theta(\boldsymbol{a}|s) \nabla_{\theta_s} \log \boldsymbol{\pi}^{\theta_s}(\boldsymbol{a}|s)^T]$$
$$= \sum_{i \in \mathcal{N}} \mathbb{E}_{a^i \sim \pi_i^{\theta_s}(a^i|s)} [\nabla_{\theta_s} \log \pi_i^{\theta_s}(a^i|s) \nabla_{\theta_s} \log \pi_i^{\theta_s}(a^i|s)^T].$$

Therefore $\mathcal{F}(\theta_s)$ is a block-diagonal matrix, and each block is corresponding to the policy parameter of an agent. Since the pseudo-inverse of a block-diagonal matrix is block-diagonal with the pseudo-inverse of each block of the original matrix, VPG has the same dynamics as global NPG.　□

## A.3 Proof of Lemma 5.7

Proof. As the gradient of the value functions equals the potential function, we prove the smoothness of value functions. Define the $\tilde{\Phi}^i(s, \boldsymbol{\pi}) = \mathbb{E}[\sum_{t=0}^\infty \gamma^t r^i(s_t, \boldsymbol{a}_t)]$, where the expectation is taken with respect to $\boldsymbol{a}_t \sim \boldsymbol{\pi}_i(\cdot|s_t)$, $s_{t+1} \sim P(\cdot|s_t, \boldsymbol{a}_t)$. The value function can be decomposed

$$\Phi(s, \boldsymbol{\pi}) = \tilde{\Phi}^i(s, \boldsymbol{\pi}) + \mathcal{H}(\boldsymbol{\pi})$$

where $\mathcal{H}(\boldsymbol{\pi}) = -\mathbb{E}[\sum_{t=0}^\infty \gamma^t \boldsymbol{\pi}(\boldsymbol{a}_t|s_t) \log \boldsymbol{\pi}(\boldsymbol{a}_t|s_t)]$. We first bound the smoothness of $\tilde{\Phi}^i(s, \boldsymbol{\pi})$. Let $\boldsymbol{\pi}^\alpha := \boldsymbol{\pi}^{\theta + \alpha u}$, where $u$ is a unit vector.

$$\left. \frac{d\boldsymbol{\pi}^\alpha(\boldsymbol{a} \mid s)}{d\alpha} \right|_{\alpha=0} = \boldsymbol{\pi}(\boldsymbol{a}|s) \sum_{i \in \mathcal{N}} \sum_{a' \in \mathcal{A}_i} u_{i, s, a', a^{-i}} (\mathbb{I}_{a' = a^i} - \pi(a'|s, a^{-i}))$$

$$\left| \left. \frac{d\boldsymbol{\pi}^\alpha(\boldsymbol{a} \mid s)}{d\alpha} \right|_{\alpha=0} \right|$$
$$= \left| \boldsymbol{\pi}(\boldsymbol{a}|s) \sum_{i \in \mathcal{N}} \sum_{a' \in \mathcal{A}_i} u_{i, s, a', a^{-i}} (\mathbb{I}_{a' = a^i} - \pi_i(\cdot|s, a^{-i})) \right|$$
$$\leq \boldsymbol{\pi}(\boldsymbol{a}|s) n \left( |u_{i, s, a^i, a^{-i}}| + \sum_{i \in \mathcal{N}} \sum_{a' \in \mathcal{A}_i} |u_{i, s, a^i, a^{-i}} \pi_i(a'|s, a^{-i})| \right)$$
$$\leq (n + 1) \boldsymbol{\pi}(\boldsymbol{a}|s)$$

$$\left. \frac{d^2 \boldsymbol{\pi}^\alpha(\boldsymbol{a} \mid s')}{(d\alpha)^2} \right|_{\alpha=0}$$
$$= \boldsymbol{\pi}(\boldsymbol{a}|s) (u_{i, s, a^i, a^{-i}} u_{j, s, a^j, a^{-j}} - \sum_{i, j \in \mathcal{N}} \sum_{a' \in \mathcal{A}_j} u_{i, s, a^i, a^{-i}} u_{j, s, a', a^{-j}} \pi_j(a'|s, a^{-j})$$
$$- \sum_{i, j \in \mathcal{N}} \sum_{a' \in \mathcal{A}_i} u_{j, s, a', a^{-i}} u_{i, s, a^j, a^{-h}} \pi_i(a'|s, a^{-i})$$
$$+ 2 \sum_{i, j \in \mathcal{N}} \sum_{a' \in \mathcal{A}_i} \sum_{a'' \in \mathcal{A}_j} u_{i, s, a', a^{-i}} u_{j, s, a'', a^{-j}} \pi_i(a'|s, a^{-i}) \pi_j(a''|s, a^{-j})$$
$$+ \sum_{i \in \mathcal{N}} \sum_{a' \in \mathcal{A}_i} u_{i, s, a', a^{-i}}^2 \pi_i(a'|s, a^{-i}))$$

$$\left| \left. \frac{d^2 \boldsymbol{\pi}^\alpha(\boldsymbol{a} \mid s')}{(d\alpha)^2} \right|_{\alpha=0} \right|$$
$$\leq \boldsymbol{\pi}(\boldsymbol{a}|s) \Bigg( u_{i, s, a^i, a^{-i}} u_{j, s, a^j, a^{-j}}$$
$$+ \Big| \sum_{i, j \in \mathcal{N}} \sum_{a' \in \mathcal{A}_j} u_{i, s, a^i, a^{-i}} u_{j, s, a', a^{-j}} \pi_j(a'|s, a^{-j}) \Big|$$
$$+ \Big| \sum_{i, j \in \mathcal{N}} \sum_{a' \in \mathcal{A}_i} u_{j, s, a', a^{-i}} u_{i, s, a^j, a^{-h}} \pi_i(a'|s, a^{-i}) \Big|$$
$$+ 2 \Big| \sum_{i, j \in \mathcal{N}} \sum_{a' \in \mathcal{A}_i} \sum_{a'' \in \mathcal{A}_j} u_{i, s, a', a^{-i}} u_{j, s, a'', a^{-j}} \pi_i(a'|s, a^{-i}) \pi_j(a''|s, a^{-j}) \Big|$$
$$+ \Big| \sum_{i \in \mathcal{N}} \sum_{a' \in \mathcal{A}_i} u_{i, s, a', a^{-i}}^2 \pi_i(a'|s, a^{-i}) \Big| \Bigg)$$
$$\leq 2(1 + n + n^2) \boldsymbol{\pi}(\boldsymbol{a}|s)$$

Let $\tilde{P}(\alpha)$ be the state-action transition matrix under $\boldsymbol{\pi}$,

$$[\tilde{P}(\alpha)]_{(s, \boldsymbol{a}) \to (s', \boldsymbol{a}')} = \boldsymbol{\pi}^\alpha(\boldsymbol{a}'|s') P(s'|s, \boldsymbol{a}).$$

We can differentiate $\tilde{P}(\alpha)$ w.r.t $\alpha$ to get

$$\left[ \left. \frac{d\tilde{P}(\alpha)}{d\alpha} \right|_{\alpha=0} \right]_{(s, \boldsymbol{a}) \to (s', \boldsymbol{a}')} = \left. \frac{d\boldsymbol{\pi}^\alpha(\boldsymbol{a}' \mid s')}{d\alpha} \right|_{\alpha=0} P(s'|s, \boldsymbol{a}).$$

For an arbitrary vector $x$,

$$\left[ \left. \frac{d\tilde{P}(\alpha)}{d\alpha} \right|_{\alpha=0} x \right]_{s, \boldsymbol{a}} = \sum_{\boldsymbol{a}', s'} \left. \frac{d\boldsymbol{\pi}^\alpha(\boldsymbol{a}' \mid s')}{d\alpha} \right|_{\alpha=0} P(s'|s, \boldsymbol{a}) x_{\boldsymbol{a}', s'}$$

$$\max_{\|u\|_2=1} \left| \left[ \frac{d\widetilde{P}(\alpha)}{d\alpha} \bigg|_{\alpha=0} x \right]_{s,\boldsymbol{a}} \right|$$

$$= \max_{\|u\|_2=1} \left| \sum_{\boldsymbol{a}',s'} \frac{d\boldsymbol{\pi}^{\alpha}(\boldsymbol{a}' \mid s')}{d\alpha} \bigg|_{\alpha=0} P(s' \mid s, \boldsymbol{a}) x_{\boldsymbol{a}',s'} \right|$$

$$\leq \sum_{\boldsymbol{a}',s'} \left| \frac{d\boldsymbol{\pi}^{\alpha}(\boldsymbol{a}' \mid s')}{d\alpha} \bigg|_{\alpha=0} \right| |P(s' \mid s, \boldsymbol{a})| |x_{\boldsymbol{a}',s'}|$$

$$\leq \sum_{s'} P(s' \mid s, \boldsymbol{a}) \|x\|_{\infty} \sum_{\boldsymbol{a}'} \left| \frac{d\boldsymbol{\pi}^{\alpha}(\boldsymbol{a}' \mid s')}{d\alpha} \bigg|_{\alpha=0} \right|$$

$$\leq \sum_{s'} P(s' \mid s, \boldsymbol{a}) \|x\|_{\infty} (n+1)$$

$$\leq (n+1)\|x\|_{\infty}.$$

By definition of $\ell_{\infty}$ norm,

$$\max_{\|u\|_2=1} \left\| \frac{d\widetilde{P}(\alpha)}{d\alpha} x \right\|_{\infty} \leq (n+1)\|x\|_{\infty}$$

Similarly, we get

$$\left[ \frac{d^2\widetilde{P}(\alpha)}{(d\alpha)^2} \bigg|_{\alpha=0} \right]_{(s,\boldsymbol{a}) \to (s',\boldsymbol{a}')} = \frac{d^2\boldsymbol{\pi}^{\alpha}(\boldsymbol{a}' \mid s')}{(d\alpha)^2} \bigg|_{\alpha=0} P(s' \mid s, \boldsymbol{a}).$$

An identical argument leads to that, for arbitrary $x$,

$$\max_{\|u\|_2=1} \left\| \frac{d^2\tilde{P}(\alpha)}{(d\alpha)^2} \bigg|_{\alpha=0} x \right\|_{\infty} \leq 2(1+n+n^2)\|x\|_{\infty}$$

$$\max_{\|u\|_2=1} \|(I-\gamma\tilde{P}(\alpha))^{-1}x\|_{\infty} = \|\sum_{n=0}^{\infty} \gamma^n \tilde{P}(\alpha)^n x\|_{\infty} \leq \frac{1}{1-\gamma}\|x\|_{\infty}$$

Denote $Q^i(s, \boldsymbol{a}; \boldsymbol{\pi}^{\alpha})$ as the action value function of $\boldsymbol{\pi}^{\alpha}$.

$$\max_{\|u\|_2=1} \left| \frac{dQ^i(s, \boldsymbol{a}; \boldsymbol{\pi}^{\alpha})}{d\alpha} \right|$$

$$= \max_{\|u\|_2=1} \gamma \left| e_{i,s,\boldsymbol{a}}^T (I-\gamma\tilde{P}(\alpha))^{-1} \frac{d\widetilde{P}(0)}{d\alpha} (I-\gamma\tilde{P}(\alpha))^{-1} r \right|$$

$$\leq \frac{\gamma(n+1)}{(1-\gamma)^2},$$

where $r$ is the reward function.

$$\max_{\|u\|_2=1} \left| \frac{d^2 Q^i(s, \boldsymbol{a}; \boldsymbol{\pi}^{\alpha})}{d\alpha^2} \right|$$

$$= \max_{\|u\|_2=1} \left| 2\gamma^2 e_{i,s,\boldsymbol{a}}^T (I-\gamma\tilde{P}(\alpha))^{-1} \frac{d\widetilde{P}(0)}{d\alpha} (I-\gamma\tilde{P}(\alpha))^{-1} \frac{d\widetilde{P}(0)}{d\alpha} (I-\gamma\tilde{P}(\alpha))^{-1} \right.$$

$$\left. + \gamma(I-\gamma\tilde{P}(\alpha))^{-1} \frac{d^2\widetilde{P}(0)}{d\alpha^2} (I-\gamma\tilde{P}(\alpha))^{-1} \right|$$

$$\leq \max_{\|u\|_2=1} \left| 2\gamma^2 e_{i,s,\boldsymbol{a}}^T (I-\gamma\tilde{P}(\alpha))^{-1} \frac{d\widetilde{P}(0)}{d\alpha} (I-\gamma\tilde{P}(\alpha))^{-1} \frac{d\widetilde{P}(0)}{d\alpha} (I-\gamma\tilde{P}(\alpha))^{-1} \right|$$

$$+ \left| \gamma(I-\gamma\tilde{P}(\alpha))^{-1} \frac{d^2\widetilde{P}(0)}{d\alpha^2} (I-\gamma\tilde{P}(\alpha))^{-1} \right|$$

$$\leq \frac{2\gamma^2(n+1)^2}{(1-\gamma)^3} + \frac{2\gamma(1+n+n^2)}{(1-\gamma)^2}$$

$$\tilde{\Phi}(s, \boldsymbol{\pi}^{\alpha}) = \sum_{\boldsymbol{a} \in \mathcal{A}} \boldsymbol{\pi}^{\alpha}(\boldsymbol{a}|s) Q^i(s, \boldsymbol{a}; \boldsymbol{\pi}^{\alpha})$$

$$\frac{d^2\tilde{\Phi}(s, \boldsymbol{\pi}^{\alpha})}{d\alpha^2} \bigg|_{\alpha=0} = \sum_{\boldsymbol{a} \in \mathcal{A}} \boldsymbol{\pi}^{\alpha}(\boldsymbol{a}|s) \frac{d^2 Q^i(s, \boldsymbol{a}; \boldsymbol{\pi}^{\alpha})}{d\alpha^2} \bigg|_{\alpha=0}$$

$$+ \sum_{\boldsymbol{a} \in \mathcal{A}} \frac{d^2 \boldsymbol{\pi}^{\alpha}(\boldsymbol{a} \mid s)}{d\alpha^2} \bigg|_{\alpha=0} Q^i(s, \boldsymbol{a}; \boldsymbol{\pi}^{\alpha})$$

$$+ 2 \sum_{\boldsymbol{a} \in \mathcal{A}} \frac{d\boldsymbol{\pi}^{\alpha}(\boldsymbol{a} \mid s)}{d\alpha} \bigg|_{\alpha=0} \frac{dQ^i(s, \boldsymbol{a}; \boldsymbol{\pi}^{\alpha})}{d\alpha} \bigg|_{\alpha=0}$$

$$\left| \frac{d^2\tilde{\Phi}(s, \boldsymbol{\pi}^{\alpha})}{d\alpha^2} \bigg|_{\alpha=0} \right| \leq \left| \sum_{\boldsymbol{a} \in \mathcal{A}} \boldsymbol{\pi}^{\alpha}(\boldsymbol{a}|s) \frac{d^2 Q^i(s, \boldsymbol{a}; \boldsymbol{\pi}^{\alpha})}{d\alpha^2} \bigg|_{\alpha=0} \right|$$

$$+ \left| \sum_{\boldsymbol{a} \in \mathcal{A}} \frac{d^2 \boldsymbol{\pi}^{\alpha}(\boldsymbol{a} \mid s)}{d\alpha^2} \bigg|_{\alpha=0} Q^i(s, \boldsymbol{a}; \boldsymbol{\pi}^{\alpha}) \right|$$

$$+ 2 \left| \sum_{\boldsymbol{a} \in \mathcal{A}} \frac{d\boldsymbol{\pi}^{\alpha}(\boldsymbol{a} \mid s)}{d\alpha} \bigg|_{\alpha=0} \right| \left| \frac{dQ^i(s, \boldsymbol{a}; \boldsymbol{\pi}^{\alpha})}{d\alpha} \bigg|_{\alpha=0} \right|$$

$$\leq \frac{2\gamma^2(n+1)^2}{(1-\gamma)^3} + \frac{2\gamma(1+n+n^2)}{(1-\gamma)^2}$$

$$+ \frac{2(1+n+n^2)}{1-\gamma} + \frac{2\gamma(n+1)^2}{(1-\gamma)^2}$$

$$< \frac{2(n+1)^2}{(1-\gamma)^3}$$

The second step is to bound the smoothness of $\mathcal{H}(\boldsymbol{\pi})$.

$$-(\boldsymbol{\pi}^{\alpha})^T \log \boldsymbol{\pi}^{\alpha} = -(\boldsymbol{\pi}^{\alpha})^T (\theta + \alpha u) + n \log \sum_{\boldsymbol{a} \in \mathcal{A}} \exp(\theta + \alpha u)$$

$$\left| -\frac{d^2 (\pi^\alpha)^T \log \pi^\alpha}{d\alpha^2} \bigg|_{\alpha=0} \right|$$

$$\leq \sum_{a \in \mathcal{A}} \left| \frac{d^2 \pi^\alpha (a \mid s)}{d\alpha^2} \bigg|_{\alpha=0} \theta \right|$$

$$+ 2 \sum_{a \in \mathcal{A}} \left| \frac{d\pi^\alpha (a \mid s)}{d\alpha} \bigg|_{\alpha=0} \right| |u|$$

$$+ n \max_{i \in \mathcal{N}} \left| 1^T \operatorname{diag}(\pi_i \odot u) - \pi_i (\pi_i \odot u)^T \right|$$

$$\leq 2(n^2 + n + 1) \frac{1 + \log \max_{i \in \mathcal{N}} |A_i|}{1 - \gamma} + 3n + 2$$

$$\left| \frac{d^2 \mathcal{H}(\pi^\alpha)}{d\alpha^2} \bigg|_{\alpha=0} \right| = \left| -\frac{d^2}{d\alpha^2} \mathbb{E}[\sum_{t=0}^{\infty} \gamma^t \pi^\alpha(a_t|s_t) \log \pi^\alpha(a_t|s_t)] \bigg|_{\alpha=0} \right|$$

$$\leq \frac{1}{1-\gamma} \left\| \frac{d^2 (\pi^\alpha)^T \log \pi^\alpha}{d\alpha^2} \right\|_\infty$$

$$\leq 2(n^2 + n + 1) \frac{1 + \log \max_{i \in \mathcal{N}} |A_i|}{(1-\gamma)^2} + \frac{3n+2}{1-\gamma} := c$$

Therefore the potential function is $(\frac{2(n+1)^2}{(1-\gamma)^3} + c)$-smooth. □

## A.4 Proof of Theorem 5.8

PROOF. Since $\Phi$ is bounded, the monotone sequence $\{\Phi^{(t)}\}_{t=0}^{\infty}$ converges to fixed point. We denote $\pi^* = \lim_{t \to \infty} \pi^{(t)}$. Assume that $\tilde{\pi}$ is a Nash equilibrium policy. We first derive the performance difference between $\tilde{\pi}$ and $\pi^*$.

$$\tilde{\Phi}(s; \tilde{\pi}) - \tilde{\Phi}(s; \pi^*) = \tilde{\Phi}(s; \tilde{\pi}) - \Phi(s; \tilde{\pi})$$
$$+ \Phi(s; \tilde{\pi}) - \Phi(s; \pi^*) + \Phi(s; \pi^*) - \tilde{\Phi}(s; \pi^*)$$
$$\leq \Phi(s; \pi^*) - \tilde{\Phi}(s; \pi^*) \leq \frac{\log |A|}{1-\gamma}$$

Note that opponent modeling will introduce extra estimation error. We denote the potential function derived by opponent modeling as $\hat{\Phi}$. And the policy derived using opponent modeling is $\hat{\pi}^*$.

$$\|\Phi(\cdot; \pi^*) - \Phi(\cdot; \hat{\pi}^*)\|_\infty = \|\Phi(\cdot; \pi^*) - \hat{\Phi}(\cdot; \pi^*) + \hat{\Phi}(\cdot; \pi^*)$$
$$- \hat{\Phi}(\cdot; \hat{\pi}^*) + \hat{\Phi}(\cdot; \hat{\pi}^*) - \Phi(\cdot; \hat{\pi}^*)\|_\infty$$
$$\leq \|\hat{Q}^i(s, a; \pi^*) - Q^i(s, a; \pi^*)\|_\infty \leq \delta$$

Therefore the performance difference is $\delta + \frac{\log |A|}{1-\gamma}$. □

## A.5 Proof of Proposition 5.9

PROOF. The objective of agent $j$ is to maximize $\log P(O_{0:\infty}^j | q)$.

$$\log P(O_{0:\infty}^j | q)$$
$$= \log \sum_{a_{0:\infty}, s_{0:\infty}} P(O_{0:\infty}^j, a_{0:\infty}, s_{0:\infty} | q)$$
$$= \log \sum_{a_{0:\infty}, s_{0:\infty}} q(a_{0:\infty}, s_{0:\infty}) \frac{P(O_{0:\infty}^j, a_{0:\infty}, s_{0:\infty} | q)}{q(a_{0:\infty}, s_{0:\infty})}$$
$$\geq \sum_{a_{0:\infty}, s_{0:\infty}} q(a_{0:\infty}, s_{0:\infty}) \log \frac{P(O_{0:\infty}^j, a_{0:\infty}, s_{0:\infty} | q)}{q(a_{0:\infty}, s_{0:\infty})}$$
$$= \mathbb{E}_{a_{0:\infty}, s_{0:\infty} \sim q} \left[ \sum_{t=0}^{\infty} r^j(s_t, a_t) - \operatorname{KL}(\rho_j(a_t^j|s_t) \| \hat{\pi}_j(a_t^j|s_t)) \right.$$
$$\left. - \operatorname{KL}(q(a_t^{-j}|s_t) \| \pi_{-j}(a_t^{-j}|s_t)) \bigg| q \right]$$
$$= \mathbb{E}_{a_0, s_0 \sim q} \left[ Q_\rho^j(s_0, a_0; \rho) - \operatorname{KL}(\rho_j(a_0^j|s_0) \| \hat{\pi}_j(a_0^j|s_0)) \bigg| q \right]$$
$$= \mathbb{E}_{s_0 \sim q} \left[ -\operatorname{KL}\left( \rho_j(a_0^j|s_0) \bigg\| \frac{\hat{\pi}_j(a_0^j|s_0) \exp(\mathbb{E}_{a_0^{-j} \sim q}[Q_\rho^j(s_0, a_0; \rho)])}{\mathbb{E}_{a_0^j \sim \hat{\pi}_j(\cdot|s_0)} \left[ \exp(\mathbb{E}_{a_0^{-j} \sim q}[Q_\rho^j(s_0, a_0; \rho)]) \right]} \right) \bigg| q \right]$$
$$+ \mathbb{E}_{s_0, a_0 \sim q} \left[ Q_\rho^j(s_0, a_0; \rho) \right]$$

From the non-negativity of KL divergence, the optimal opponent model of agent $j$ is

$$\rho_j^*(a_0^j|s_0) = \frac{\hat{\pi}_j(a_0^j|s_0) \exp(\mathbb{E}_{a_0^{-j} \sim q}[Q_\rho^j(s_0, a_0; \rho)])}{\mathbb{E}_{a_0^j \sim \hat{\pi}_j(\cdot|s_0)} \left[ \exp(\mathbb{E}_{a_0^{-j} \sim q}[Q_\rho^j(s_0, a_0; \rho)]) \right]}$$

□